\documentclass[aps,prl,showpacs,twocolumn,floatfix]{revtex4-1}
\usepackage[utf8x]{inputenc}
\usepackage{textcomp}
\usepackage{amsmath,amssymb,latexsym,graphicx}
\usepackage{mathtools}
\usepackage{overpic}
\usepackage{color}
\usepackage{subfigure}
\usepackage{natbib}
\usepackage[colorlinks=true,a4paper=true, pdfstartview=FitV,linkcolor=blue, citecolor=blue, urlcolor=blue]{hyperref}
\usepackage[all]{hypcap}
\usepackage{times}
\usepackage{multirow}
\usepackage{bm}
\usepackage{mathrsfs}
\usepackage{relsize}
\newcommand*{\diff}{\mathop{}\!\mathrm{d}}

\def\blackdot{$\mbox{\fontsize{14}{1}\selectfont $\bullet$}$}
\def\redsquare{${\color{red}\blacksquare}$}

\def\magentadiamond{$\color{magenta}\diamondsuit$}

\begin{document}
\title{Field-induced $p$-wave Superconductivity in Mesoscopic Systems}
\author{Jia-Wei Huo$^1$, Wei-Qiang Chen$^{2,1}$, S. Raghu$^3$, Fu-Chun Zhang$^{1,4}$}
\email{fuchun@hku.hk}
\affiliation{$^1$Department of Physics and Center of Theoretical and Computational Physics, The University of Hong Kong, Hong Kong, China\\
$^2$Department of Physics, South University of Science and Technology of China, Shenzhen, Guangdong 518055, China\\
$^3$Department of Physics, Stanford University, Stanford, CA, 94305\\
$^4$Department of Physics, Zhejiang University, Hangzhou, China}
\date{\today}

\begin{abstract}
By using Bogoliubov-de Gennes equations, we study superconducting (SC) states in a quasi 2-dimensional system of radius $R$. It is shown that no vortices exist in $s$-wave SC samples with $R<R_\text{c}\sim\xi(0)$, the $T=0$ coherence length. We predict that chiral $p$-wave states exhibit superconductivity for $R<R_\text{c}$ only in the presence of a vortex with opposite chirality. This {\it induced} SC phase is a consequence of non-zero chirality of the pairing order parameter and implies the presence of chiral edge currents. Our study may be applied to sharply probing the pairing symmetry of unconventional superconductors.
\end{abstract}
\maketitle

One of the most fascinating areas in condensed matter physics is the study of unconventional superconductors that spontaneously break time reversal symmetry \cite{Sigrist1991}, such as $p+ip$ and $d+id$ superconductors. Due to their non-trivial topology, such systems may support exotic objects such as half-quantum vortices and zero energy modes of Majorana quasi-particles \cite{Ivanov2001}. A key challenge is to identify the pairing symmetry of such superconductors. Currently, the Kerr effect and muon spin relaxation experiments have played an important role in detecting broken time-reversal symmetry.

Here we propose that mesoscopic systems may provide novel platforms for the study of unconventional superconductivity due to the experimental availability and intriguing properties in small length scale. Recent advancements in nano-fabrication have made it possible to study superconductivity in mesoscopic samples with size comparable to the coherence length $\xi$ \cite{Geim1996}. As the dimension of a superconductor is reduced to order of the SC coherence length, the effect due to the edge and the vortex core becomes important. The former enforces vanishing quasi-particle amplitudes, while the latter requires a topological phase factor on the pairing potential, whose amplitude vanishes at the center of the vortex core. All these have provided motivation for the research of few-vortex physics \cite{Chibotaru2000,*Moshchalkov1995,*Fink1985,*Mertelj2003,*Benoist1997,*Schweigert1998a}.

In this letter, we use BdG equations to study SC states in a disk of radius $R$, comparable to the coherence length $\xi$. This microscopic theory allows to study the system at zero temperature and with size smaller than $\xi$ quantitatively, and also enables us to evaluate observables in detail. Our results are further supported by intuitive analysis based on the (Ginzberg-Landau) GL theory. To be consistent with previous studies based on the GL theory \cite{Schweigert1998}, we begin with $s$-wave superconductors, and confirm that the vortex state can only be generated with system size larger than a critical value $R_{\text{c}}$. For the unconventional paring, we predict that chiral $p$-wave states exhibit superconductivity for $R<R_{\text{c}}$ only in the presence of a vortex with opposite chirality, namely an {\it induced} SC phenomenon. A half-quantum vortex in the equal spin pairing $p$-wave state is also studied. The relevance to the possible $p$-wave superconductor $\text{Sr}_2\text{RuO}_4$ is discussed.

We start with the $s$-wave case. The BdG equations read
\begin{eqnarray}\label{eqn:bdg}
\left[\begin{array}{cc}
  h_0(\bm{r})&\Delta(\bm{r})\\
  \Delta^*(\bm{r})&-h_0^*(\bm{r})
\end{array}\right]\left[\begin{array}{c}
  u_i(\bm{r})\\
  v_i(\bm{r})
\end{array}\right]=E_i\left[\begin{array}{c}
  u_i(\bm{r})\\
  v_i(\bm{r})
\end{array}\right].
\end{eqnarray}
In Eq.\ref{eqn:bdg}, $u_i(\mathbf{r})$ and $v_i(\mathbf{r})$ form the
two-component wavefunction of quasi-particles corresponding to
energy $E_i$ in the SC state. $\Delta(\mathbf{r})$ is the pairing
potential satisfying the self-consistent equation
\begin{equation}
  \Delta(\bm{r})=g\!\!\sum_{E_i<\Lambda}\!\!u_i(\bm{r})v^*_i(\bm{r})[1-2f(E_i)],
\end{equation}
with $g$ the coupling constant, $f(E_i)$ the Fermi distribution
function, and $\Lambda$ an energy cut-off. $h_0(\mathbf{r})$ is the
single electron Hamiltonian
\begin{align}
\label{eq:3}
h_0(\bm{r})&=\frac{1}{2m}\left[-i\hbar \nabla-\frac{e}{c}\bm{A}(\bm{r})
\right]^2-\mu,
\end{align}
where $m$ is the electron mass, $\mu$ is the chemical potential. $\bm{A}(\bm{r})$ is the vector potential determined by the Maxwell's equation $\nabla\times\nabla\times\bm{A}=\frac{4\pi}{c}\bm{j}$, with the supercurrent density
\begin{multline}
\label{eq:2}
\bm{j}(\bm{r})=\frac{e\hbar}{2mi}\sum_{i}\left\{f(E_i)u^*_i(\bm{r})\left[\nabla-\frac{ie}{\hbar c}\bm{A}(\bm{r})\right]u_i(\bm{r})+\right.\\
\left.[1-f(E_i)]v_i(\bm{r})\left[\nabla-\frac{ie}{\hbar c}\bm{A}(\bm{r})\right]v_i^*(\bm{r})-\text{H.c.}\right\}.
\end{multline}
We work in a polar coordinate system $(r,\theta)$, and consider a natural
boundary condition $u_i(r\!=\!R,\theta)=v_i(\!r=\!R,\theta)=0$. Note that we neglect the $z$-dependence on $u$, $v$, and $\bm{A}$ due to the fact that effective mass along $z$-direction is large and the thickness of the system is small enough \cite{Tanaka2002}. The single
particle wavefunction of $h_0$ (with $\bm{A}=0$) corresponding to the eigenvalue $\epsilon_{jl}$ is
\begin{equation}
  \phi_{j,l}(r,\theta)=\frac{\sqrt{2}}{RJ_{l+1}(\alpha_{jl})}
  J_l\left(\alpha_{jl}\frac{r}{R}\right) e^{i l \theta},
\end{equation}
where $l$ is the angular momentum, $J_l(x)$ is the $l^{\text{th}}$
order Bessel function of the first kind, and $\alpha_{jl}$ is the
$j^{\text{th}}$ zero of $J_l(x)$ \cite{Gygi1991}. In the presence
of pairing, we have $u_i(\mathbf{r})\!=\!u_i(r) e^{il\theta}$ and $v_i(\mathbf{r})\!=\!v_i(r) e^{i(l-n)\theta}$,
with $n$ the vorticity. The order parameter then takes the form
$\Delta(\mathbf{r})\! =\! \Delta(r) e^{in\theta}$. $n\!=\!0$ corresponds
to a vortex-free state, and $n\!=\!1$ to a vortex state. $u(r)$
and $v(r)$ can be expanded in terms of $\phi_{j,l}$. The
coherence length may be estimated $\xi(T)\!=\!\hbar v_{\text{F}}/\pi\Delta(T)$ \cite{Gennes1999},
with $v_{\text{F}}$ the Fermi velocity and $\Delta$ the order parameter of the
vortex-free state at $R \rightarrow \infty$.

\begin{figure}[ht]\centering
  \begin{tabular}{cc}
    \resizebox{.25\textwidth}{!}{
      \begin{overpic}{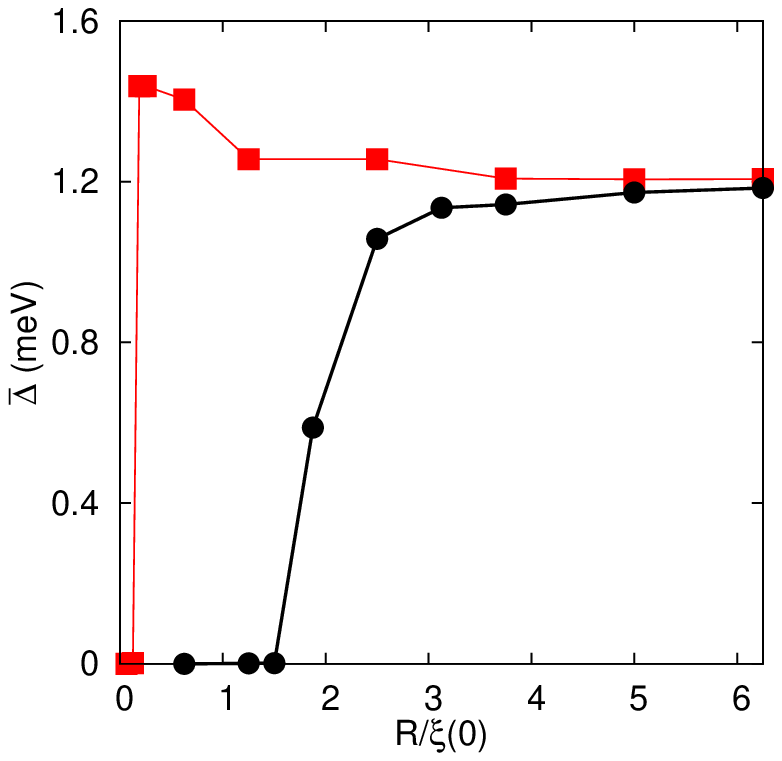}
        \put(0,87){\textbf{\large{(a)}}}
      \end{overpic}} &
    \resizebox{.25\textwidth}{!}{
      \begin{overpic}{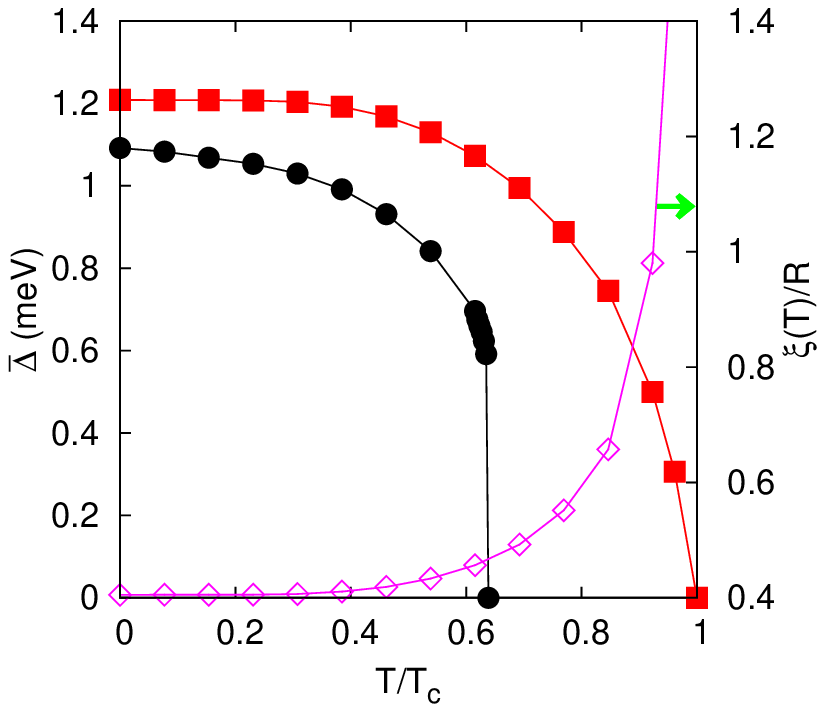}
        \put(0,87){\textbf{\large{(b)}}}
      \end{overpic}}
  \end{tabular}
  \resizebox{.3\textwidth}{!}{
    \begin{overpic}{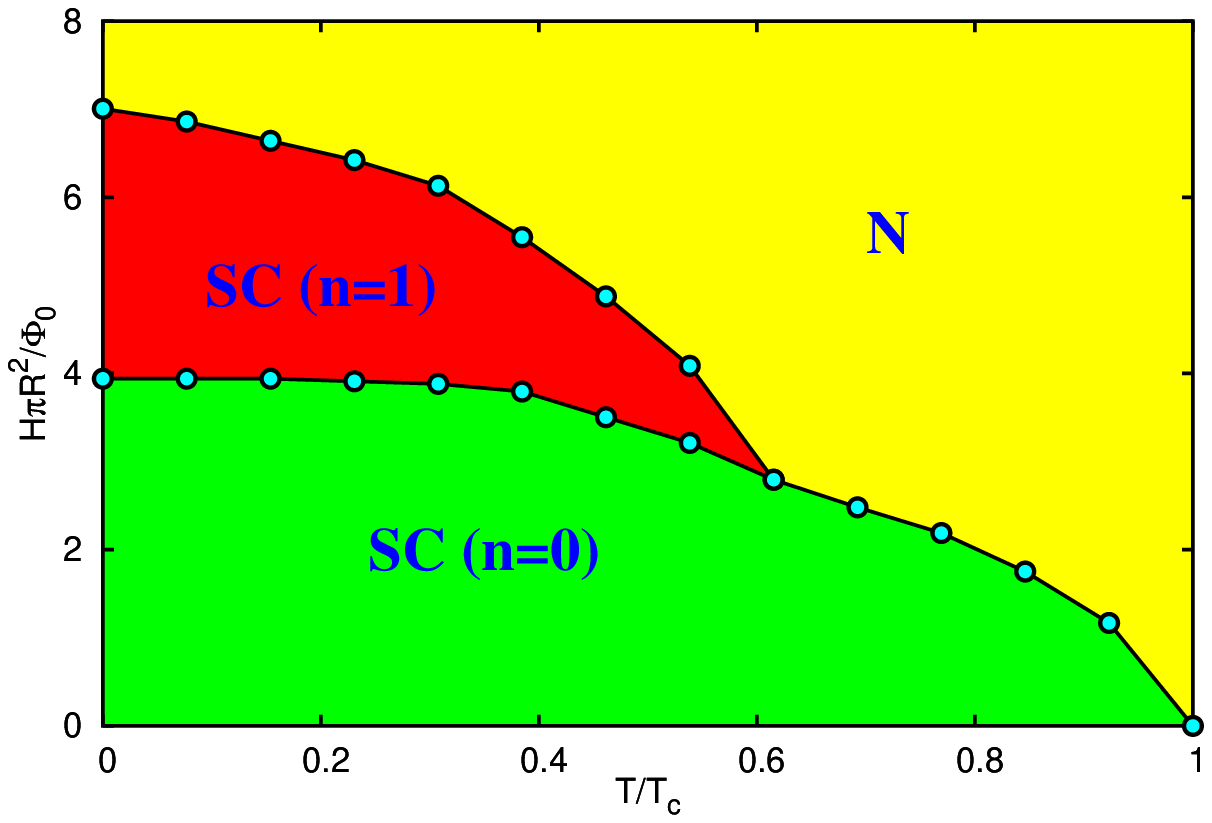}
      \put(-2,65){\textbf{\Large{(c)}}}
    \end{overpic}}
  \caption{\label{s_wave} (color online) $s$-wave superconductivity
 in a disk of radius $R$. (a) Spatially averaged SC gap $\bar{\Delta}$ vs. $R$ at $T\!=\!0$. (b) $\bar{\Delta}$ and the coherence length $\xi$ (\magentadiamond) vs. $T$ for $R\!=\!2.5\xi(0)$. {\redsquare} is for the vortex free state, and {\blackdot} for the vortex state in a magnetic field. (c) Phase diagram $H$ vs. $T$ for $R\!=\!2.5\xi(0)$. Here $\Phi_0$ is the superconducting flux quantum. The parameters are $gm/2\pi\hbar^2\!=\!0.256$, $\mu\!=\!180$ meV, $\Lambda\!=\!30$ meV and $m\!=\!m_\text{e}$, the free electron mass, which give the SC gap $\Delta=1.2$ meV, and $\xi(0)\!\approx\!40$ nm in the large $R$ limit.}
\end{figure}

In Fig.~\ref{s_wave} (a) and (b), we plot the spatially averaged SC order parameters $\bar\Delta$. In Fig.~\ref{s_wave} (a), for the vortex-free state ($n=0$) at $T=0$, the superconductivity remains robust as $R$ decreases. The sudden drop in $\bar{\Delta}$ at $R\ll\xi(0)$ is due to the quantum size effect, where the energy level spacing due to confinement becomes comparable to the SC gap \cite{Anderson1959}. The vortex state ($n=1$) vanishes at $R<R_\text{c}\approx 1.5 \xi(0)$, consistent with the GL theory \cite{Schweigert1998} and the recent experiment \cite{Nishio2008}. Therefore, there is a sharp crossover between vortex and vortex-less behaviors at $R\approx 1.5 \xi(0)$.

At small $R\gtrsim R_\text{c}$, the vortex state exists at low $T$,
but vanishes at $T= T'< T_\text{c}$, due to the increasing
vortex core size (or the value of $\xi(T)$) with $T$, as shown in
Fig.~\ref{s_wave} (b). The corresponding phase diagram in Fig.~\ref{s_wave} (c) can be calculated by comparing the Gibbs free energy $\mathcal{G}=\langle\mathcal{H}\rangle-T\mathcal{S}+\mathcal{F}_H$ of different states in equilibrium \cite{Kosztin1998}, with $\mathcal{H}$ the mean-field Hamiltonian, $\mathcal{S}$ the entropy and $\mathcal{F}_H=\int\diff\bm{r}\frac{|\bm{B}-\bm{H}|^2}{8\pi}$ the magnetic field exclusion energy \footnote{The magnetic induction $\bm{B}=\nabla\times \bm{A}$, is a sum of the external uniform magnetic field $\bm{H}$ and the field induced by the supercurrent. In our calculations, by increasing $\bm{H}$, we can change $\bm{A}$ to find the critical $\bm{H}$ values of the SC states.}. There is a sharp drop for the upper critical field at $T\approx 0.6T_\text{c}$, indicating the suppression of the vortex state. This result may explain the recent experiment on ultra small aluminum thin films, showing a similar drop below $T_{\text{c}}$ in the $H-T$ phase diagram \cite{Staley2011}.

We now study the much more intriguing $p$-wave case. We shall only consider states which are allowed by rotational invariance and shall restrict our attention to the chiral $p_x+ip_y$ phase, the 2D analog of $^3$He-A \cite{Rice1995}. For a given spin component, the BdG equations have been derived by Matsumoto $et$ $al.$ \cite{Matsumoto1999,*Matsumoto2001} and read
\begin{eqnarray}
\label{eq:p_wave}
\left[\begin{array}{cc}
  h_0(\bm{r})&\Pi(\bm{r})\\
  -\Pi^*(\bm{r})&-h_0^*(\bm{r})
\end{array}\right]\left[\begin{array}{c}
  u_i(\bm{r})\\
  v_i(\bm{r})
\end{array}\right]=E_i\left[\begin{array}{c}
  u_i(\bm{r})\\
  v_i(\bm{r})
\end{array}\right],
\end{eqnarray}
with $\Pi(\bm{r})=-\frac{i}{k_\text{F}}\sum_\pm\left[\Delta_\pm\square_\pm+
\frac{1}{2}(\square_\pm\Delta_\pm)\right]$. Here $\Delta_\pm$ are
the pairings of $p_x\pm ip_y$ states, respectively, and satisfy
the self-consistent equations
\begin{multline}
\Delta_\pm(\bm{r})=-i\frac{g}{2k_\text{F}}\!\!\sum_{E_i<\Lambda}\!\!\left[v_i^*(\bm{r})\square_\mp u_i(\bm{r})-\right. \\
\left. u_i(\bm{r})\square_\mp v_i^*(\bm{r})\right][1-2f(E_i)],
\end{multline}
with $k_\text{F}=\sqrt{2m\mu/\hbar^2}$ and $\square_\pm=e^{\pm
i\theta}(\partial_r\pm\frac{i}{r}\partial_\theta)$ \footnote{The off-diagonal terms in Eq.~\ref{eq:p_wave} are fully gauge invariant under the simultaneous phase transformation of the quasi-particle amplitudes and the superconducting order parameters.}. In the thermodynamic limit, there are two degenerate eigenstates with
$p_x\pm ip_y$ pairings. In the finite system, the two pairing
symmetries are mixed near the boundaries. We consider here the state
where the $p_x +ip_y$ pairing is the dominant component. In the disk
geometry, the pairing parameters have the forms,
$\Delta_+(\bm{r})=\Delta_+(r)e^{i n\theta}$ and
$\Delta_-(\bm{r})=\Delta_-(r)e^{i(n+2)\theta}$, where $n$ is the vorticity.
We focus on three cases: the vortex-free state ($n=0$),
negative-vortex state ($n=-1$), and positive-vortex state
($n=1$).
\begin{figure}[ht]\centering
    \begin{tabular}{cc}
      \resizebox{.25\textwidth}{!}{
        \begin{overpic}{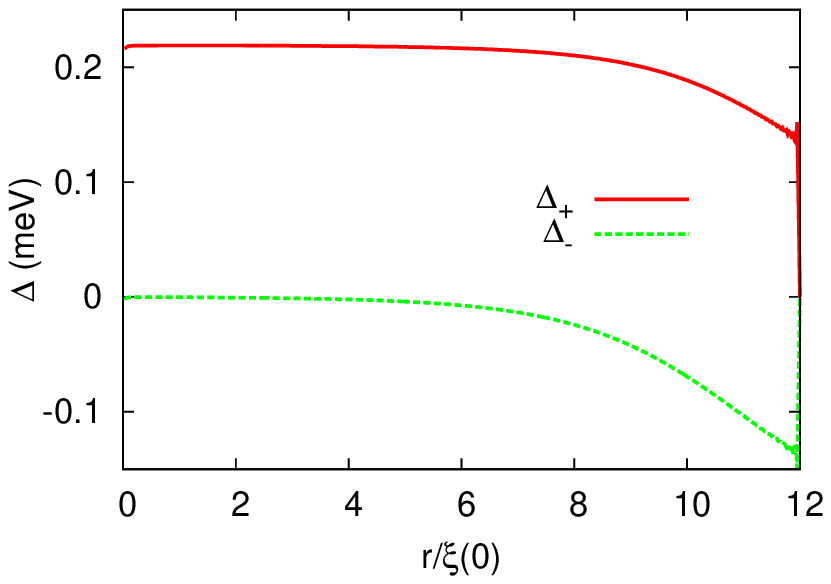}
          \put(0,65){\textbf{\large{(a)}}}
        \end{overpic}} &
      \resizebox{.25\textwidth}{!}{
        \begin{overpic}{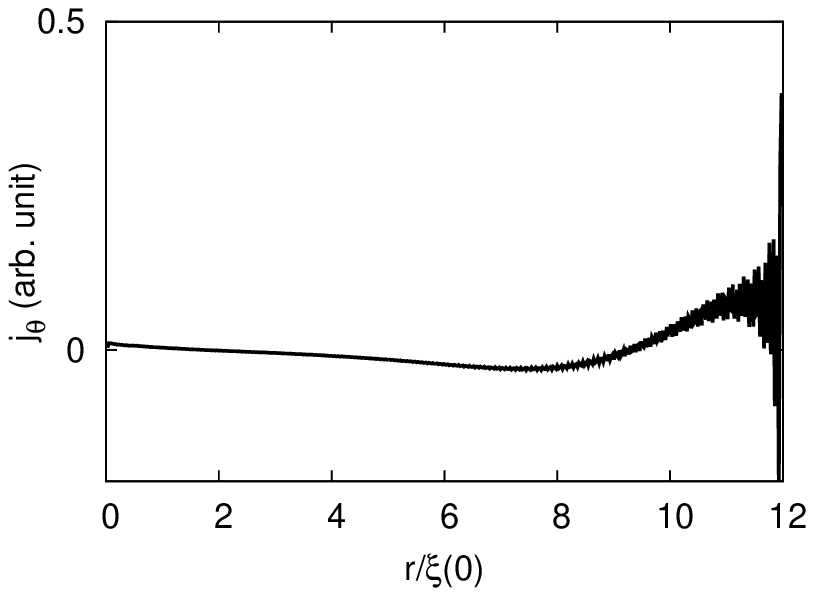}
          \put(0,65){\textbf{\large{(b)}}}
          \put(28,28){\includegraphics[width=0.18\textwidth]{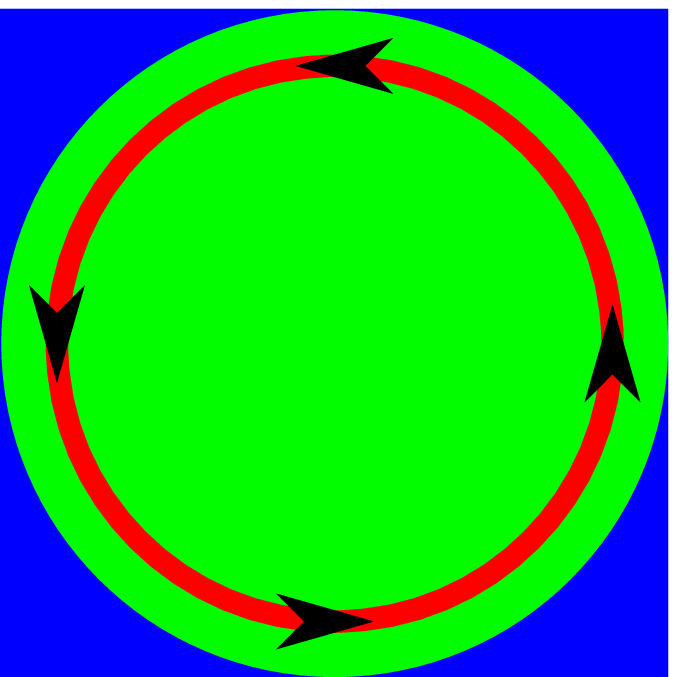}}
        \end{overpic}} \\
      \resizebox{.25\textwidth}{!}{
        \begin{overpic}{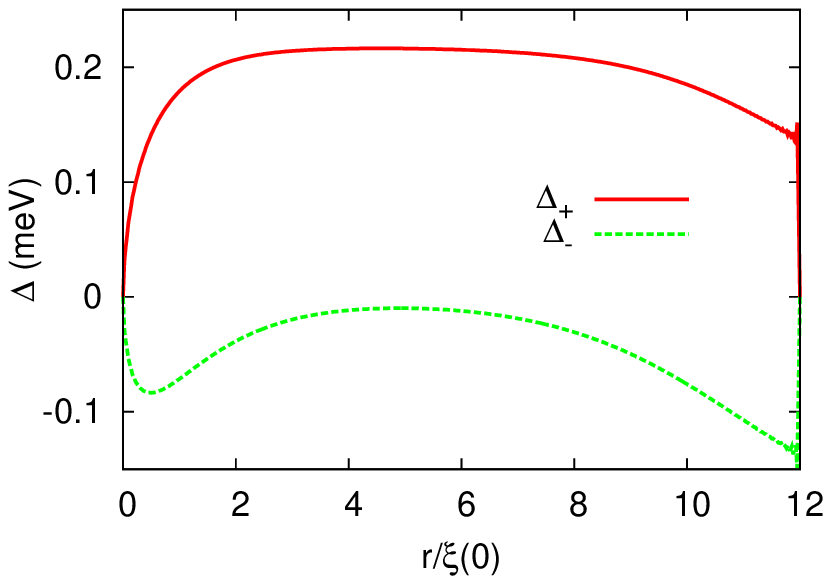}
          \put(0,65){\textbf{\large{(c)}}}
        \end{overpic}} &
      \resizebox{.25\textwidth}{!}{
        \begin{overpic}{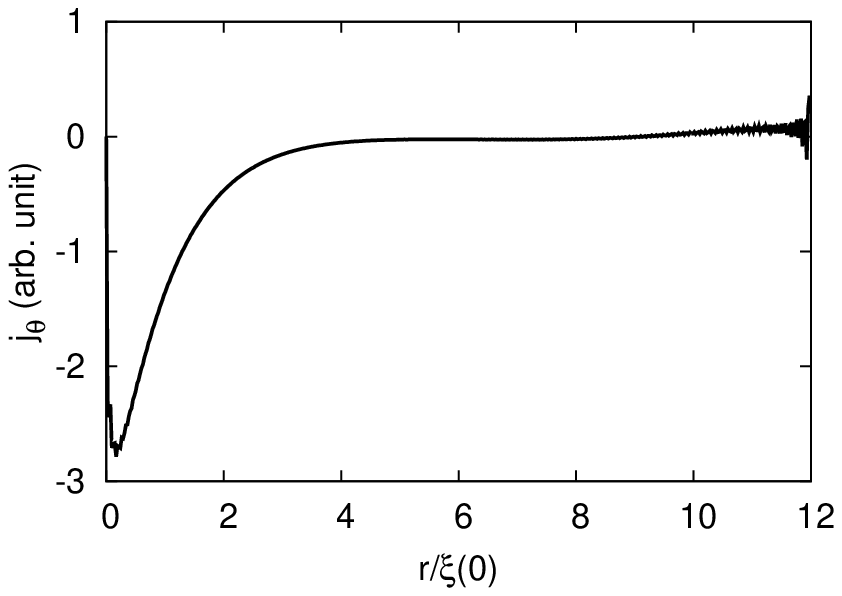}
          \put(0,65){\textbf{\large{(d)}}}
          \put(50,14){\includegraphics[width=0.18\textwidth]{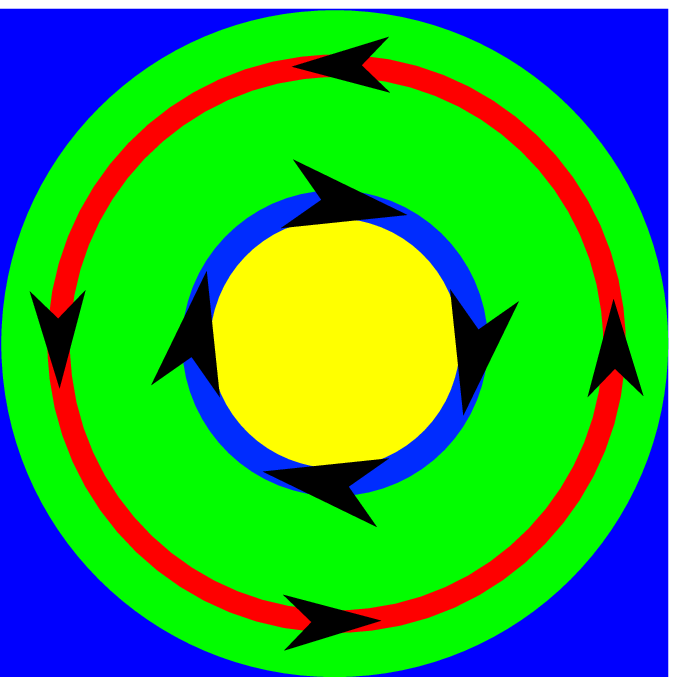}}
        \end{overpic}} \\
      \resizebox{.25\textwidth}{!}{
        \begin{overpic}{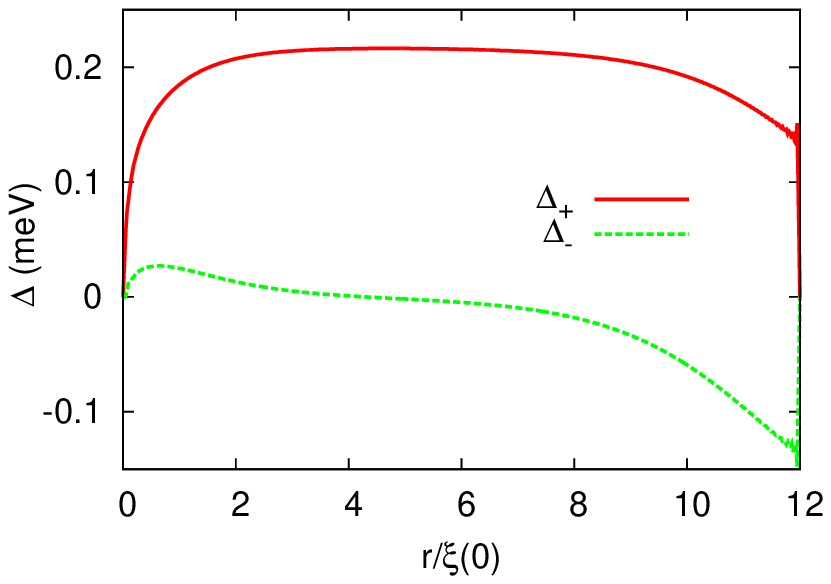}
          \put(0,65){\textbf{\large{(e)}}}
        \end{overpic}} &
      \resizebox{.25\textwidth}{!}{
        \begin{overpic}{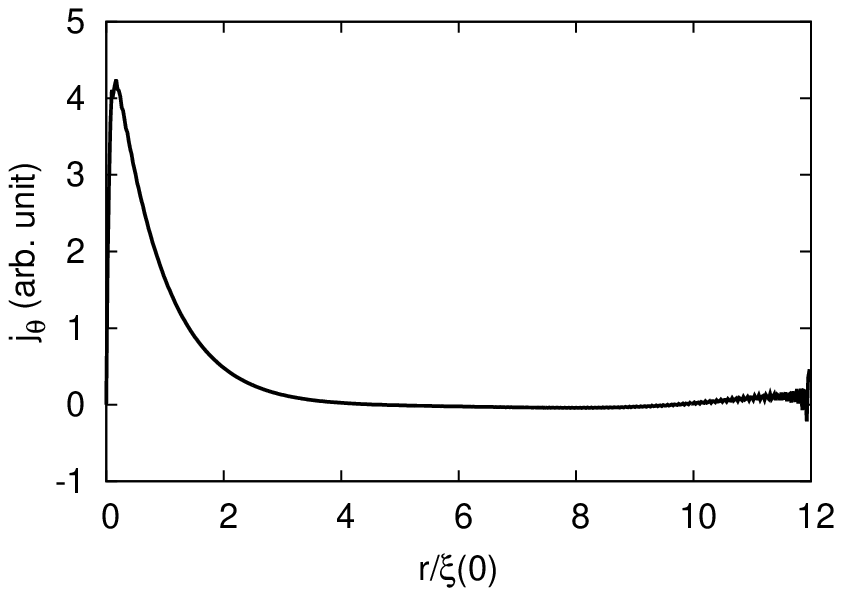}
          \put(0,65){\textbf{\large{(f)}}}
          \put(50,25){\includegraphics[width=0.18\textwidth]{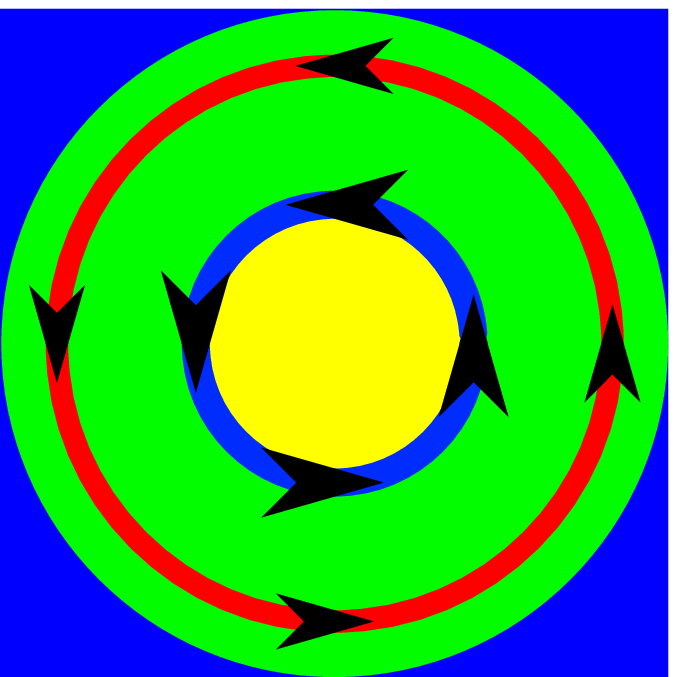}}
        \end{overpic}}
    \end{tabular}
    \caption{\label{p_wave_size}(Color online)
    Spin triplet $p_x \pm ip_y$ state ($s_z\!=\!0$) in a disk of radius $R\!=\!12\xi(0)$.
    Left: Spatial variation of the order parameters $\Delta_+$ and
    $\Delta_-$ (imaginary part).
    Right: Spatial dependence of the supercurrent density $j_\theta(r)$.
    Upper panels (a), (b): vortex-free state;
    middle panels (c), (d): negative-vortex state;
    and lower panels (e), (f): positive-vortex state.
    The insets on the right panels illustrate the corresponding current
    flowing directions at the edge and near the vortex core.
    The parameters are $gm/2\pi\hbar^2\!=\!0.2$, $\mu\!=\!\Lambda\!=\!16.32$ meV and $m\!=\!m_\text{e}$,
    which give $T_\text{c}$=1.4 K, and $\xi(T\!=\!0)\!\approx\!70$ nm, close to those of $\text{Sr}_2\text{RuO}_4$ \cite{Mackenzie2003}.}
\end{figure}
Let us consider the spin triplet case with $s_z=0$. Our results
are shown in Fig.~\ref{p_wave_size}. For the vortex-free state,
the pairing is dominated by the $p_x+ip_y$ component $\Delta_+$,
except near the edge where $p_x-ip_y$ component $\Delta_-$ also becomes
substantial. At the edge $r$=$R$, $|\Delta_+|\approx|\Delta_-|$. The right panels in
Fig.~\ref{p_wave_size} show the density distributions of the supercurrents.
Note that at the edge, the current density shows rapid oscillations
with a wave vector $2k_\text{F}$. In Fig.~\ref{p_wave_size} (b),
near the edge within the length scale $\xi(0)$ \cite{Stone2004}, the supercurrent flows counterclockwise, while next to
the edge current within the penetration depth, about $2\xi(0)$, a weak screening current flowing clockwise is driven by the Meissner effect. We also confirm that the bound edge states contribute to the former and the scattering states contribute to the latter \cite{Iniotakis2008,*Zare2010}.

Due to the broken time reversal symmetry, there are two types of
vortices with $n=\pm1$, depending on the direction of the applied
magnetic field. Note that there is an induced supercurrent near
the vortex core, whose direction is the same as that of the intrinsic
chiral edge currents for $n=1$ and is opposite to that for $n=-1$.

\begin{figure}[ht]\centering
  \begin{tabular}{cc}
    \resizebox{.25\textwidth}{!}{
      \begin{overpic}{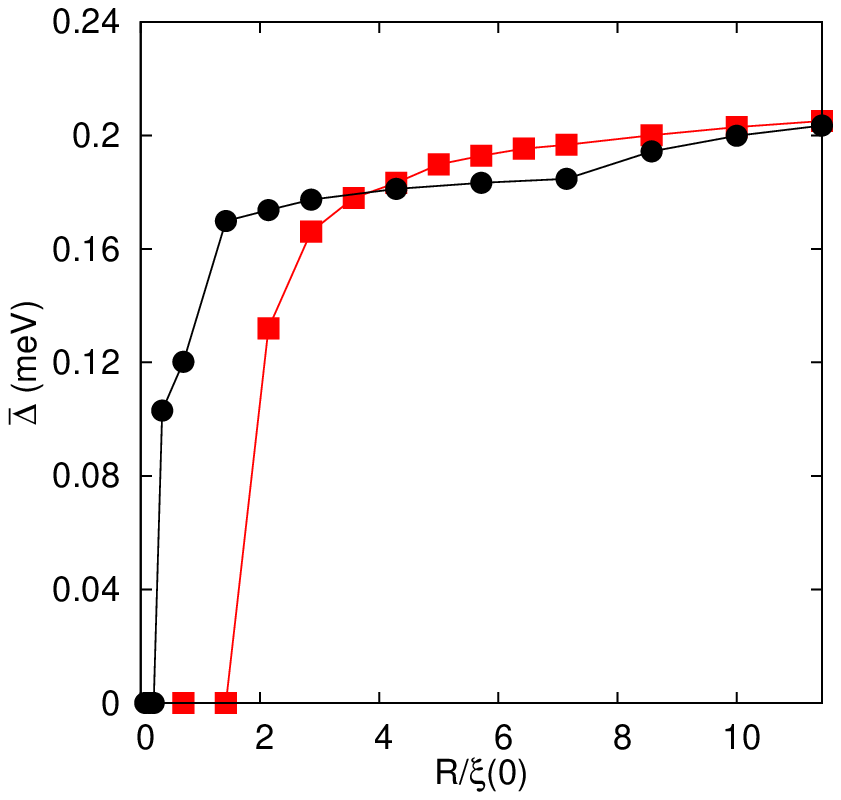}
        \put(-4,95){\textbf{\Large{(a)}}}
      \end{overpic}} &
    \resizebox{.25\textwidth}{!}{
      \begin{overpic}{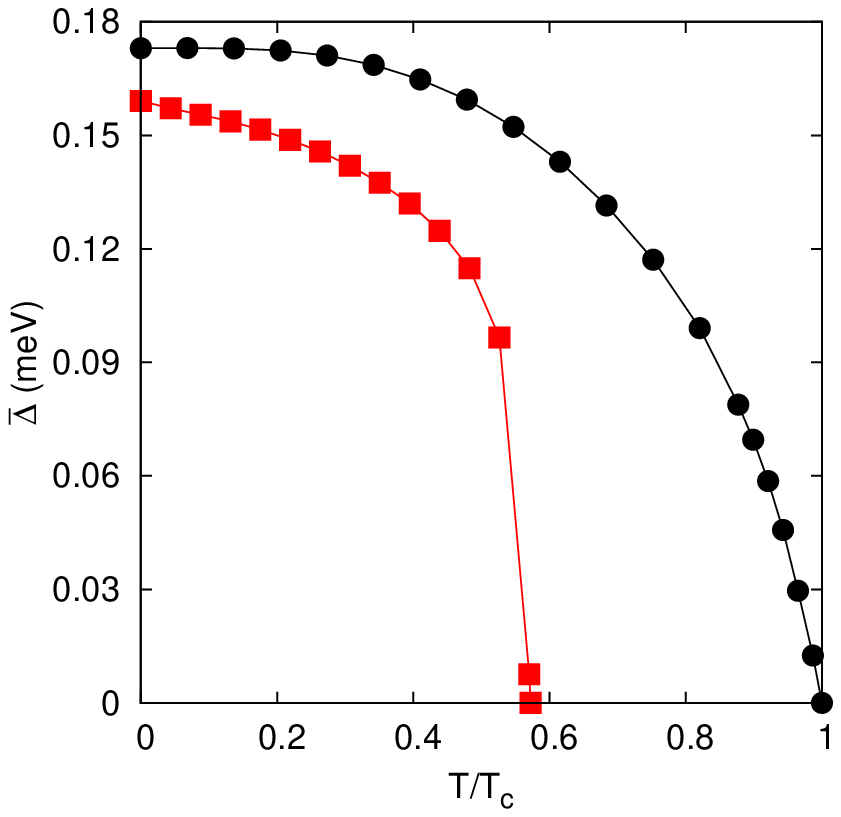}
        \put(-4,95){\textbf{\Large{(b)}}}
      \end{overpic}} \\
    \resizebox{.25\textwidth}{!}{
      \begin{overpic}{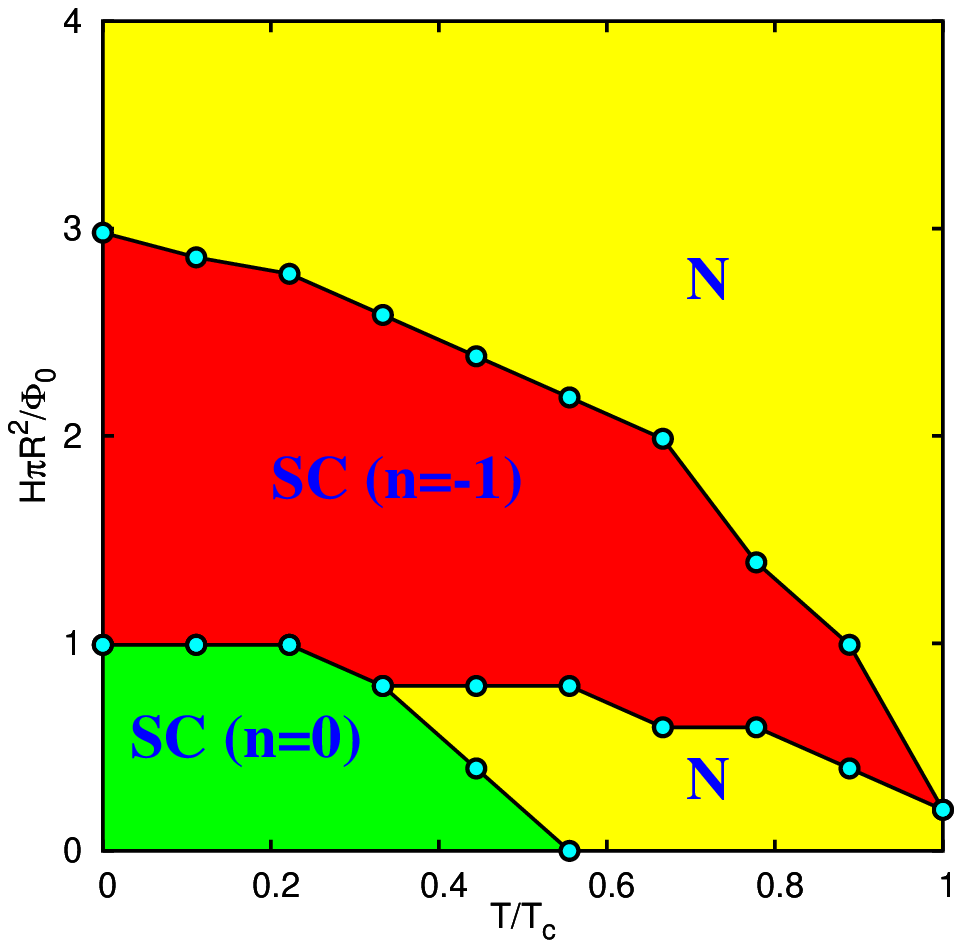}
        \put(-4,95){\textbf{\Large{(c)}}}
      \end{overpic}} &
    \resizebox{.25\textwidth}{!}{
      \begin{overpic}{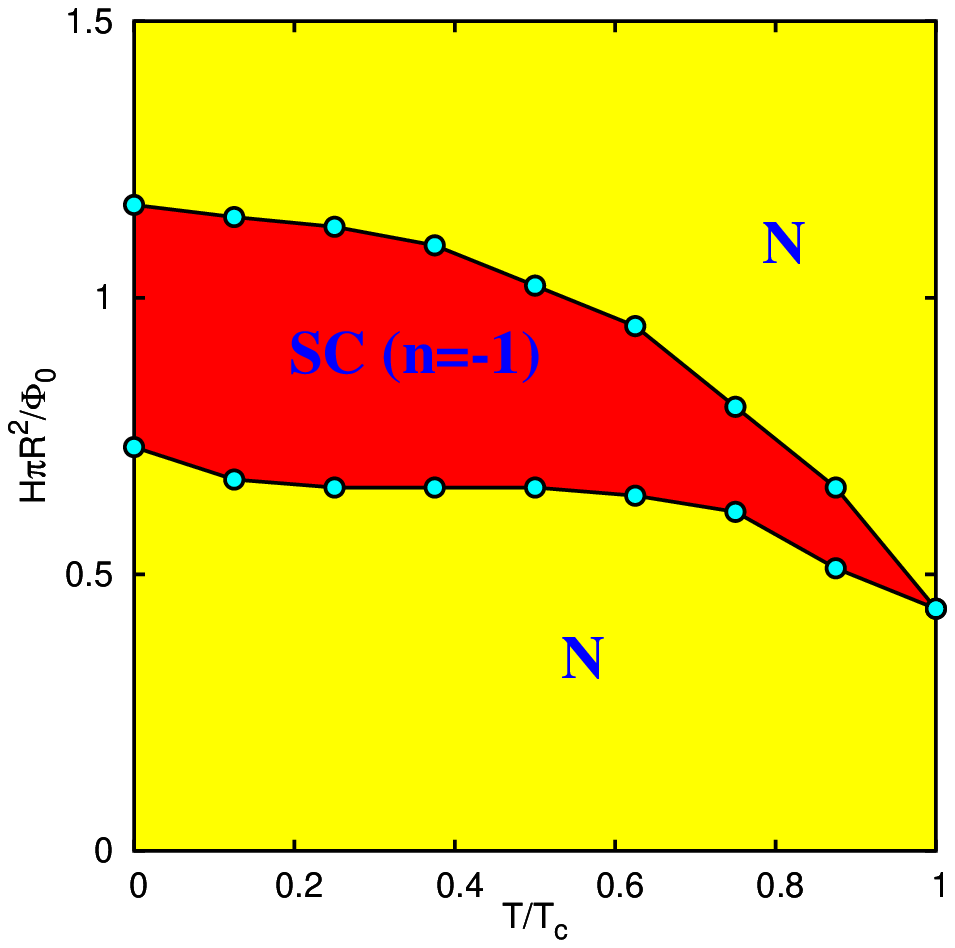}
        \put(-4,95){\textbf{\Large{(d)}}}
      \end{overpic}}
  \end{tabular}
  \caption{\label{p_wave} (Color online) Spatially averaged order parameter $\Delta\!=\!\sqrt{|\Delta_+|^2+|\Delta_-|^2}$ as functions of disk radius $R$ at $T$=0 (panel a), and  as functions of $T$ for a given $R\!=\!2.4\xi(0)$ (panel b), in a spin triplet $p_x \pm ip_y$-wave superconductor with $s_z\!=\!0$. {\redsquare} is for the vortex-free state, and {\blackdot} is for the negative-vortex state in a magnetic field. The parameters are the same as in Fig.~\ref{p_wave_size}. Panels (c) and (d) are phase diagrams for $R=2.4\xi(0)\gtrsim R_\text{c}$ and $R=1.4\xi(0)<R_\text{c}$, respectively. In both cases, the upper critical fields are close to $\Phi_0/2\pi\xi(T)^2$.}
\end{figure}

Now we turn to the study of the quantum size effect due to finite
confinement at $T=0$. Contrary to the $s$-wave SC state, there is
no vortex-free $p$-wave SC state at $R<R_\text{c}\approx1.5\xi(0)$, and
the negative-vortex state in the $p$-wave state is robust against
small size, as we can see from Fig.~\ref{p_wave} (a). The
superconductivity in the vortex-free state is destroyed by the
supercurrent induced by the chiral motion of the Cooper pairs at
the edge, which becomes dominant in a sufficiently small sample.
In fact, the boundary region in the vortex-free $p$-wave
SC state is similar to the vortex core in the $s$-wave
state where the superconductivity is suppressed by the
supercurrent around the vortex core. In the negative-vortex
state, the supercurrents around the vortex and around boundary
flow in the opposite directions (see Fig.~\ref{p_wave_size} (d)),
and partially cancel each other, which makes the
superconductivity stable. This situation is similar to that of a vortex-antivortex pair \cite{Mertelj2003}.

To propose experiments to test our findings, below we study the $T$-dependence of SC states for the system of $R=2.4\xi(0)$. As can be seen in Fig.~\ref{p_wave} (b), the $T$-dependence of the order parameters for different SC states show qualitative discrepancy. In the absence of any external magnetic field, the chiral $p$-wave superconductivity disappears above $0.6T_{\text{c}}$. This is because that $\xi(T)$ increases with temperature. If a magnetic field is applied to the system to induce a negative-vortex, the SC condensate may revive between $0.6T_{\text{c}}$ and $T_{\text{c}}$. To further illustrate the whole physical picture quantitatively, we can construct the corresponding phase diagram in Fig.~\ref{p_wave} (c) by comparing the Gibbs free energy $\mathcal{G}$ as usual. It is clear that this phase diagram and its $s$-wave counterpart in Fig.~\ref{s_wave} (c) form a sharp contrast. It shows both {\it re-entrant} and {\it induced} $p$-wave SC phases in various regions, both of which are absent in the $s$-wave counterpart. Specifically, between $0.3T_\text{c}$ and $0.6T_\text{c}$, there is a {\it re-entrant} SC phenomenon as $H$ increases from zero since it evolves between two different SC states by crossing a normal phase; however, the SC($n$=-1) state becomes {\it induced} from $0.6T_\text{c}$ to $T_\text{c}$ due to the absence of vortex-free state. Similarly, a phase diagram for $R=1.4\xi(0)<R_\text{c}$ is presented in Fig.~\ref{p_wave} (d), showing the complete disappearance of the vortex-free state and the {\it re-entrant} phenomenon. Thus in this case, field-cooled samples with $H\sim\Phi_0/\pi R^2$ will exhibit superconductivity whereas zero-field cooled samples do not. This prediction can be readily tested by using Sr$_2$RuO$_4$ microcrystals \cite{Note_Liu,*Cai2012}. Although strong evidence suggests that the SC state of this material has odd parity \cite{Nelson2004}, the null result on the observation of the edge currents \cite{Bjornsson2005,*Kirtley2007} seems to shed doubt on its chiral $p$-wave symmetry \cite{Rice1995}. Therefore, it will be exciting to confirm our prediction of the {\it induced} superconductivity in a magnetic field, since it can provide a very strong evidence of the chiral $p$-wave SC pairing for Sr$_2$RuO$_4$.

To understand the finite size effects in both the $s$-wave and
$p$-wave pairings from the topological point of view, we can
define a winding number associated with the SC order parameter as
\begin{equation}\label{eqn:winding}
  \mathcal{W}=\frac{1}{2\pi i}\oint_\mathcal{C}\frac{\diff\Delta}{\Delta},
\end{equation}
where $\Delta$ refers to the off-diagonal part of the BdG
equations and $\mathcal{C}$ is a circle around the origin counterclockwise. For the conventional $s$-wave superconductors, the winding number $\mathcal{W}$ is just the vorticity $n$.
However, for the $p$-wave pairing with $p_x+ip_y$ as the dominant
component, we find that $\mathcal{W}=n+1$, {\it i.e.}, the sum of
the vorticity and the additional $+1$ for the intrinsic chirality.
Therefore, for the $p$-wave vortex-free state the winding number
is 1, similar to the vortex $s$-wave state. These two states with
$\mathcal{W}=1$ are found to vanish below a critical size
comparable to the coherence length. On the contrary, both the $s$-wave vortex-free state
and the $p$-wave negative-vortex state with $\mathcal{W}=0$ survive even below the critical size \footnote{We speculate that the time-reversal invariant ``B'' phase vortex-free state (analogous to the B phase in $^3$He) also lies in this category.}. These results are in good agreement with an intuitive analysis on the GL free-energy density. We only focus on the dominant gradient term $|\bm{\mathcal{D}}\Delta|^2$ \cite{Gennes1999},
where the covariant derivative $\bm{\mathcal{D}}$ reads
\begin{equation}
\bm{\mathcal{D}}\equiv-i\nabla-2\bm{A}=-i\partial_r\hat{\bm{e}}_r-(i\frac{1}{r}\partial_\theta+2A_\theta)\hat{\bm{e}}_\theta
\end{equation}
in a rotationally invariant system, and we set $\hbar$=$e$=$c$=$1$.
Here $\frac{1}{r}\partial_\theta\Delta$ dominates and makes the SC state disfavored, as the size is reduced.
To avoid this energetically unfavorable term, $\partial_\theta\Delta$ has to vanish, {\it i.e.}, $\mathcal{W}=0$, leading to the survival of the $s$-wave vortex-free state and the $p$-wave negative-vortex state.

The above result may be extended to discuss the equal-spin pairing
(ESP) phase in a spin triplet $p$-wave superconductor, which may
support half-quantum vortices \cite{Jang2011}. In this phase, there
are two weakly interacting condensates with Cooper-pair spin
configurations $|\!\!\uparrow\uparrow\rangle$ and
$|\!\!\downarrow\downarrow\rangle$ coupled by the electromagnetic
field. The Cooper pair wave function is given by \cite{Ivanov2001}
\begin{multline}
  \Psi(\bm{r})\propto(\Delta_{\uparrow+}(r)e^{\pm i\theta}|\!\uparrow\uparrow\rangle+\Delta_{\downarrow+}(r)|\!\downarrow\downarrow\rangle)*(p_x+ip_y)\\
  +e^{2i\theta}(\Delta_{\uparrow-}(r)e^{\pm i\theta}|\!\uparrow\uparrow\rangle+\Delta_{\downarrow-}(r)|\!\downarrow\downarrow\rangle)*(p_x-ip_y),
\end{multline}
where $\Delta_{\sigma}$ denotes the pairing in the state $|\sigma\sigma\rangle$ and $*$ is the symmetrized product. Then the two condensates are described by two separated sets of self-consistent equations with different vorticities but the same vector potential. In this situation, electrons with up spin form a vortex state with $n_\uparrow=\pm1$, while those with down spin form a vortex-free state. Based on the relation between the winding number and the finite size confinement discussed previously, it is straightforward to predict that when the size of the system is reduced below $R_\text{c}$, the pairing in $|\!\!\downarrow\downarrow\rangle$ breaks down, while the pairing in $|\!\!\uparrow\uparrow\rangle$ with $n_\uparrow=-1$ is still robust. In this sense, a spinless chiral $p$-wave superconductor is achieved. We have confirmed this scenario by performing numerical simulations. We stress that the results above are based on the assumption that Cooper pairs with distinct spins are weakly coupled.

In summary, we have used BdG equations to study the finite size effect, for both $s$-wave and chiral $p$-wave superconductors. For the $p$-wave pairing, the vortex-free SC state does not exist below a critical size, whereas the vortex state is robust even for the system size as small as the coherence length, where the opposite winding of the vortex compensates the $p$-wave intrinsic winding at the boundary. These results predict a magnetic-field-induced superconductivity in ultra small samples with $p$-wave pairing such as $\text{Sr}_2\text{RuO}_4$. For all these quantum geometrical constrains for both $s$-wave and $p$-wave pairings, the winding number $\mathcal{W}$ plays a determining role. Although our mean-field theory has neglected SC fluctuations, we expect that our results should be valid for BCS superconductors such as Sr$_2$RuO$_4$ with a large coherence length and superfluid density.
\begin{acknowledgments}
We thank Y. Liu for stimulating discussions on the experiment \cite{Staley2011}. This work is partially supported by Hong Kong RGC Grant No. GRF HKU707211, GRF HKU701010, HKUST3/CRF/09 and start-up funds at Stanford University (S.R.). We acknowledge the hospitalities from the Aspen Center for physics (S.R.) and KITP Santa Barbara (F.C.Z.).
\end{acknowledgments}

\bibliography{mybib}
\bibliographystyle{apsrev4-1}
\end{document}